\documentclass[11pt]{article}
\usepackage{amssymb,cite,epsf,graphics,graphicx}
\newcommand\eq{\begin{equation}}
\newcommand\en{\end{equation}}
\newcommand\bea{\begin{eqnarray}}
\newcommand\eea{\end{eqnarray}}
\newcommand\ba{\(\begin{array}}
\newcommand\ea{\end{array}\)}
\newcommand\nn{\nonumber}

\newcommand{\ZZ}{{\hbox{$\sf\textstyle Z\kern-0.4em Z$}}}
\newcommand{\ZN}{\ZZ_N}
\newcommand{\cY}{{\cal Y}}
\newcommand{\ds}{\displaystyle}

\begin{document}

\begin{titlepage}
\vskip 0.5cm
\begin{flushright}
DCPT-03/29 \\
\end{flushright}
\vskip 1.5cm
\begin{center}
{\Large{\bf
Thermodynamic Bethe ansatz for the  AII $\sigma$-models
}}
\end{center}
\vskip 0.8cm
\centerline{Andrei Babichenko\footnote{e-mail: {\tt ababichenko@hotmail.com }}
and Roberto Tateo\footnote{e-mail: {\tt roberto.tateo@durham.ac.uk}; Fax:
0044-0191-3343085}$^,\hspace{-1mm}$
\footnote{e-mail: {\tt tateo@infn.to.it}}
}
\vskip 0.9cm
\centerline{${}^{1}$\sl\small Shahal str., 73/4, Jerusalem, 93721, Israel \,}
\centerline{${}^{2}$\sl\small Dept.~of Mathematical Sciences,
University of Durham, Durham DH1 3LE, UK\,}
\centerline{${}^{3}$\sl\small Dipartimento di Fisica Teorica,
Universit\`a di Torino, Via P. Giuria 1, 10125 Torino, I\,}
\vskip 1.25cm
\begin{abstract}

We derive thermodynamic 
Bethe ansatz equations  describing  the vacuum  energy 
of the  
$SU(2N)/Sp(N)$ nonlinear  sigma model on a cylinder geometry.
The starting points are the  recently-proposed 
amplitudes for the scattering  among the physical, massive  
excitations of the theory. 
The analysis fully confirms the correctness of the S-matrix.
We also derive  closed sets of functional relations for the pseudoenergies
(Y-systems).
These relations are shown  to be the $k \rightarrow \infty$  limit
of the $Sp(k+1)$-related systems  studied
some years ago by Kuniba and  Nakanishi
in the framework of lattice models.\\

\noindent
PACS: 02.30.Ik; 11.25.Hf; 11.10.-z \\
\noindent
Keywords: Nonlinear sigma models; Integrable models; 
Finite-size scaling; Thermodynamic Bethe ansatz
\end{abstract}
\end{titlepage}

\section{Introduction}
Since the early days of modern particle physics, the interest
in four dimensional   non-Abelian gauge theories has also strongly 
encouraged the  study of  two dimensional nonlinear sigma models.
The main, well-known reason is that the two theories    
share the remarkable properties  of 
renormalizability, confinement and  asymptotic freedom. 
Besides the obvious advantage of  working  in a lower dimensional 
setup, many 2D nonlinear sigma models exhibit the  even more useful
property of integrability. Exact solubility offers 
the  unique opportunity to investigate  non perturbative  phenomena in quantum field theory and  statistical mechanics  via non perturbative tools 
like the exact S-matrix~\cite{ZZ0}, the Form Factor~\cite{FF} and the 
thermodynamic Bethe ansatz~\cite{Za} (TBA) methods.
Along the lines of  the  earlier work by  Fateev and Zamolodchikov~\cite{FZ} 
but also motivated by new, interesting applications in condensed matter physics
and string theory~\cite{MM}  there has 
been recently  a renewed interest  in 2D integrable 
sigma models~\cite{Fsig,F1,F2,SK1,BH,BW,My,MKa} . 

When dealing with an integrable quantum field theory, a well established
strategy  is to derive the scattering matrix by solving suitably imposed 
Yang-Baxter equations,  together with  the unitary and crossing constraints. 
The S-matrix, via the so-called bootstrap procedure,  gives access to 
the full infrared mass  spectrum of the theory and
via the  Form Factors, to 
the large distance behaviour of correlation functions. 
 
In order to explore the  ultraviolet phase of the model 
and, at the same time, to  check the correctness of the  
S-matrix, the standard  tool is  the thermodynamic Bethe ansatz.
The TBA method is susceptible to all the details of the scattering matrix, 
it uses  the full mass spectrum and it is sensitive to the
ambiguous (CDD) scalar factors contributing to the amplitudes. 
It leads to numerically solvable integral equations for 
the vacuum  energy on a cylinder geometry and  exact values for 
central charge and scaling dimensions of the 
conformal field theory (CFT) describing the model  in its  deep ultraviolet limit. 

Following the idea of  Zamolodchikov~\cite{Zam2}, and  starting from a 
given 
TBA equation,
it is possible to derive   sets of   functional relations called Y-systems. 
The Y-systems  are ``universal'' in the sense 
that  there   is a one--to--many correspondence with TBAs.
A Y-system can be  transformed by imposing different  analytic 
properties, to  physically distinct TBAs.
(The relationship  between  Y-systems  and     TBA  equations 
is practically  the same  as  the one  between   differential equations  
and their solutions.)

Many   examples of functional relations  which 
exhibit  elegant 
Lie algebraic-related  structures  are now known. Most of them are,
or were at some stage,  
purely conjectures 
(see, for example, \cite{Zb,Zd,KN,KM,Ma,Ra,RTV,QRT,HolTBA,T,DPT}). 
An important  family   of  such systems   is encoded on a pair 
of Dynkin diagrams of ADE type~\cite{Ra,RTV,QRT}; a second family   
is in  correspondence with all the simple Lie algebras\footnote{These two families of Y-systems are partially overlapping.} and it 
 was studied in the works
by Kuniba and Nakanishi~\cite{KN} in the closely related framework of 
integrable lattice models.
Many, but by no means all,  solutions 
of these systems  have been successfully
related  to perturbations of minimal  CFTs and to  sigma 
models~\cite{Za,FZ,Fsig,F2,Zd,KN,KM,Ma,Ra,RTV,QRT,HolTBA,T,DPT,CFKS,DM}.

In this paper, such an  example  of  TBA system is derived 
in the context of integrable nonlinear sigma models.
Being infinite in one direction, the equations
require a truncation for the  ultraviolet
analysis. The truncated systems are found to  
describe the perturbation  of the coset
$SU(2N)_{k+1}/Sp(N)_{k+1}$ via the relevant operator  $(1,adj)$.

The starting point is  the recently proposed \cite{My} 
S-matrix  for  the sigma models defined on the
symmetric space AII. (This space is   isomorphic to the group coset manifold 
$SU(2N)/Sp(N)$.) The  S-matrix is   of   $SU(2N)$ Gross-Neveu type, 
with CDD factors 
modified in such a way that their poles describe only the  particle multiplets
related  to the even fundamental weights of $SU(2N)$. 

The  plan of the paper is as follows. The model  is briefly introduced  in  
section~\ref{S1}. 
The derivation of the  TBA equations is sketched in section~\ref{S2}. 
The Y-system is derived in section~\ref{S3} and  
its equivalence   to a particular limit of the   
$C_{k+1}$-related\footnote{To avoid any confusion: by  
$Sp(n)$ we mean   the compact real form for the classical  Lie algebra $C_n$.}
system of \cite{KN}  is discussed in section~\ref{Sn}.
Using  the dilogarithm identities of \cite{KN} 
the  ultraviolet  central charge is found to be $c=2N^{2}-N-1$:
a result which is fully  consistent  with the interpretation of 
the model as a perturbation of the $SU(2N)/Sp(N)$ conformal field theory. 
Section~\ref{S4}  contains  our conclusions, 
while the explicit definitions of the 
kernel functions are collected in the  appendix at the end of the paper. 
\section{The AII sigma models}
\label{S1}
The classical action of the quantum field theory in question is defined 
through the  functional 
\eq
S[g]= {1 \over \lambda^2} 
\int d^2x Tr[\partial_\mu g(x) \partial_\mu g^{-1}(x)]~,
\label{S0}
\en
where $g(x)$ belongs to a matrix representation of $SU(2N)$  
constrained on  the symmetric  space $SU(2N)/Sp(N)$. Then the action $S_{NLSM}[\phi] \equiv S[g(\phi)]$
of the model is obtained from $S[g]$  by 
solving the constraint in terms 
of the  independent set of fields $\{ \phi\}$  and  substituting the  
resulting  $g(\phi(x))$ into (\ref{S0}).

The  spectrum of the AII sigma models consists in multiplets transforming
in the representations with even fundamental highest weights $\mu _{2a}$ of
$SU(2N)$. The masses of these multiplets are 
\eq
m_{2a}= M \sin \left( \frac{\pi a}{N} \right),~~~~ a=1,...,N-1~, 
\en
($M=M(\lambda)$ is the mass scale). The amplitude $S_{2a,2b}$ for
scattering of a particle  $2a$  on the particle  $2b$ is  obtained 
by fusion of $S_{2,2}$. The latter was proposed  in \cite{My}, and  has the 
form \cite{HollowS}
\eq
S_{2a,2b}(\theta )=X_{2a,2b}(\theta )\sigma _{2a,2b}(\theta )\widetilde{S}%
_{2a,2b}(\theta )~~.  
\label{sm}
\en
$\widetilde{S}_{2a,2b}(\theta )$ is the $SU(2N)$ invariant R-matrix
spectral decomposed into its irreducible representations~\cite{ORW},
 $\sigma_{2a,2b}(\theta )$  is the cross-unitarity factor
and   $X_{2a,2b}(\theta )$  is again  obtained  by  fusion
of $X_{2,2}(\theta )$. 
They have the following forms \cite{ORW,HollowS}
\bea
X_{2a,2b}(\theta ) &=&\prod_{i=1}^{a}\prod_{j=1}^{b}X(\theta +4\pi i\Delta
(i+j-1-\frac{a+b}{2})) \nn \\
&=&\exp \left\{ -\int_{-\infty }^{\infty }\frac{d\omega }{\omega
}e^{i\frac{g%
}{\pi }\theta \omega }\left( \widehat{A}_{2a,2b}^{(2N)}(\omega )-\delta
_{ab}\right) \right\}~,
\eea
with  $g=2N$, $\Delta =\frac{1}{g}$, and 
\eq
 X(\theta )=
\frac{\sinh ((\theta +4\pi i\Delta )/2)}{\sinh ((\theta -4\pi i\Delta )/2)}~,
~~~~~
\sigma _{2a,2b}(\theta )=\exp \left\{ \int_{-\infty }^{\infty }\frac{d\omega
}{\omega }e^{i\frac{g}{\pi }\theta \omega }\widehat{\Psi}%
_{2a,2b}(\omega )\right\}~.
\en
The explicit expressions  of 
$\widehat{A}_{2a,2b}^{(2N)}(\omega )$ and  $\widehat{\Psi}_{2a,2b}(\omega)$
are  given in the  appendix at the end of the paper.
(Notice  that the factor $X_{2a,2b}(\theta)$ is the same as
$X_{a b}(\theta )$ in the $SU(2N)$ invariant Gross-Neveu  model of~\cite{ORW}, 
however the $\widehat{A}_{2a,2b}^{(2N)}(\omega )$
are  different. This is  because  an N-dependent form of the
Fourier transform is used in this paper.)
\noindent
Using these expressions we find
\eq
\sigma _{2a,2b}(\theta )X_{2a,2b}(\theta ) =\exp \left\{ -\int_{-\infty
}^{\infty }\frac{d\omega }{\omega }e^{i\frac{g}{\pi }\theta \omega }\left(
\widehat{Y}_{2a,2b}(\omega )-\delta _{ab}\right) \right\}~,
\en
\eq
\widehat{Y}_{2a,2b}(\omega ) =\frac{e^{|\omega |}}{\cosh \omega }\frac{
\sinh (2|\omega |\min (a,b))\sinh (2\omega (N-\max (a,b)))}{\sinh (2\omega
N)
}~.
\en

\section{Diagonalisation of the transfer matrix and the TBA equations}
\label{S2}
The standard derivation of the  TBA equations starts by imposing 
 periodicity  conditions on the wavefunction describing  a gas  of 
particles with rapidity set $\{ \theta _{l} \}$ and  
confined on a ring of circumference $\cal{L}$.
The quantisation  condition  is
\eq
e^{i m_{2a_j} { \cal L } \sinh \theta_{j}}~ T^{2a_j}(\theta_j |\theta
_{j+1},...,\theta_{\cal{N}},\theta_{1},...,\theta _{j-1})=1~,
\label{BAET}
\en
$T^{2a_j}$ is the transfer matrix and it is constructed out of the S-matrix as 
\eq
T^{2a_j}(\theta_j |\theta
_{j+1},...,\theta _{\cal{N}},\theta _{1},...,\theta _{j-1})=
-\mbox{Tr}
\left[ 
S_{2a_j, 2a_1}(\theta_j-\theta_1) \dots S_{2a_j, 2a_{\cal N}}(\theta_j-\theta_{\cal N})
\right]~,
\label{Tr}
\en
where the trace  is over the representation of the particle of type $2a_j$.
The main problem is  then to  find the  largest eigenvalue 
$T_0$  of $T$. The solution is~\cite{BR}
\begin{equation}
T_{0}^{2a_j}=\prod_{i=1,i\neq j}^{\cal{N}}
X_{2a_j,2a_{i}}(\theta_j -\theta_{i})~
\sigma _{2a_j,2a_{i}}(\theta_j -\theta _{i})\times 
\prod_{\alpha=1}^{M_{2a_j}}   
\left(
\frac{\theta_j -v_{\alpha }^{(2a_j)}+i\pi \Delta }{\theta_j -v_{\alpha
}^{(2a_j)}-i\pi \Delta } 
\right)~.   
\label{Teig}
\end{equation}%
In equations (\ref{Teig}), the set of  parameters $\{ v_{\alpha }^{(b)} \}$ 
is  solution of the Bethe equation
\cite{ORW,BR}
\begin{equation}
\prod_{j=1}^{\cal{N}} \left(\frac{v_{\alpha }^{(b)}-\theta _{j}+i\pi \Delta
\omega _{a_{j}}\cdot \alpha _{b}}{v_{\alpha }^{(b)}-\theta _{j}+i\pi \Delta
\omega _{a_{j}}\cdot \alpha _{b}} \right)=
\Omega _{\alpha
}^{(b)}\prod_{c=1}^{2N-1}\prod_{\beta =1}^{M_{c}} \left(\frac{v_{\alpha
}^{(b)}-v_{\beta }^{(c)}+i\pi \Delta \alpha _{b}\cdot \alpha _{c}}{v_{\alpha
}^{(b)}-v_{\beta }^{(c)}-i\pi \Delta \alpha _{b}\cdot \alpha _{c}}
\right)~.
\label{BE}
\end{equation}%
$\omega _{b}$ and $\alpha _{b}$ are, respectively,  the fundamental 
weights and roots of $SU(2N)$, $M_{c}$ is  the number of solutions of type 
$c$, and $\Omega_{\alpha }^{(b)}$ is a  phase factor which is 
irrelevant for the current  calculations. 
In the  thermodynamic limit  
\eq
\cal{N}\rightarrow \infty~,~~~
\cal{L}\rightarrow \infty~,~~~ \cal{N}/\cal{L}=\hbox{finite}~,
\en 
the solutions of  (\ref{BE})
group into complex ``strings'' of  length $p$ and  real part 
$v_{\alpha p}^{(b)}$
\eq
 v_{\alpha }^{(b)}=v_{\alpha p}^{(b)}+i\pi \Delta (p+1-2k),~~k=1,2,...,p~.
\en
Using   (\ref{BE}) in  equation  (\ref{BAET}) with $T$ replaced by  
$T_{0}$, introducing  
``magnon'' densities $\rho_{p}^{b}(v)$ for the real parts of 
$v_{\alpha}^{(b)}$,  densities
$\sigma_{2a}(\theta)$ for the particles, and after   
writing  the  log  of equations (\ref{BAET}) and (\ref{BE}) in terms of  
the  densities we obtain
\eq
\sigma _{2a}(\theta )+\widetilde{\sigma }_{2a}(\theta ) =\frac{m_{2a}}{%
2\pi }\cosh \theta +\sum_{b=1}^{N-1}Y_{2a,2b}\ast \sigma _{2b}(\theta
)-\sum_{p=1}^{\infty }a_{p}\ast \rho _{p}^{2a}(\theta )~,~~(a=1,\dots,N-1) 
\label{eq1}
\en
\eq
a_{p}\ast \sigma _{c}(\theta ) = \widetilde{\rho }_{p}^{c}+\sum_{q=1}^{%
\infty }\sum_{b=1}^{2N-1}A_{pq}^{(\infty )}\ast K_{cb}^{(2N)}\ast \rho
_{q}^{b}(\theta )~,~~~(c=1,\dots,2N-1)~.
\label{eq2}
\en
(The symbol $\ast $ denotes  the standard convolution.)
In (\ref{eq2} ) $\widetilde{\sigma
}_{2a}$ and $\widetilde{%
\rho }_{p}^{a}$ complement the densities of particles to the total
densities of states, and are called  densities of holes. The expressions for 
the
kernels $A_{pq}^{(\infty )}$, $K_{ab}^{(N)}$ and $a_{p}$,
together  with some useful identities, are written  in the  appendix.

Solving the latter equation with respect to  $\rho _{q}^{b}$ and plugging
the solution into equation (\ref{eq1}):
\begin{eqnarray}
\widetilde{s}_{0}^{2a}(\theta ) &=&\frac{m_{2a}}{2\pi }\cosh \theta
-\sum_{b=1}^{N-1}\sum_{q=0}^{\infty } Z_{0q}^{(\infty )}\ast
A_{2a,2b}^{(2N)}\ast s_{q}^{2b}(\theta )~,~~~(a=1,\dots,N-1)  \\
\widetilde{s}_{p}^{c}(\theta ) &=&-\sum_{b=1}^{2N-1}\sum_{q=0}^{\infty
}K_{pq}^{(\infty )}\ast A_{cb}^{(2N)}\ast s_{q}^{b}(\theta )~,~~
(c=1,\dots,2N-1)
\label{TBAlast}
\end{eqnarray}%
where $p=1,2,\dots$,
\eq
s_{0}^{c}=\sigma _{0}^{c}~,~~~~~ \widetilde{s}_{0}^{c}=\widetilde{\sigma }%
_{0}^{c}~,~~~~~ s_{p}^{c}=\widetilde{\rho }_{p}^{c}~,~~~~~\widetilde{s}_{p}^{c}=\rho
_{p}^{c}~,
\en
and 
\eq
\sigma _{0}^{2i+1}=\widetilde{\sigma }_{0}^{2i+1}=0~.
\en 
The new kernel $Z_{0q}^{(\infty )}$ has
the following Fourier representation
\eq
\widehat{Z}_{0q}^{(\infty )}(\omega )=\widehat{K}_{0q}^{(\infty
)}(\omega )\left( 1-\frac{\delta _{q0}}{2\cosh ^{2}(\omega )}\right)~.
\en
{}From equation  (\ref{TBAlast}) we see that the  excitations $\widetilde{s}%
_{p}^{2a}$ with $p=0$ are massive, whereas those with $p\geq 1$  
are of  magnonic type.
\noindent
The next step in the derivation is  also  standard:
one looks for the  minimum of the free energy $F=E-TS$ subject to the 
 constraint  (\ref{TBAlast}).( $T$ is the 
temperature, 
$E$ and $S$ are, respectively, the energy and the entropy and  are
expressible in terms of the  densities $s,\widetilde{s}$.) The final 
equations are
\begin{equation}
L_{0+}^{2a}(\theta )=R m_{2a}\cosh \theta +\sum_{b=1}^{N-1}\sum_{q=0}^{\infty
}A_{2a,2b}^{(2N)}\ast Z_{0q}^{(\infty )}\ast L_{q-}^{2b}(\theta
)~,~~~(a=1,\dots,N-1)
\label{Tba0}
\end{equation}%
\begin{equation}
L_{p+}^{c}(\theta )=\sum_{b=1}^{N-1}\sum_{q=0}^{\infty }A_{cb}^{(2N)}\ast
K_{pq}^{(\infty )}\ast L_{q-}^{b}(\theta )~,~~(c=1,\dots,2N-1)  \label{Tba2}
\end{equation}%
where $L_{p\pm }^{a}=\ln \left( 1+e^{\pm \varepsilon _{p}^{a}}\right) $, $R=1/T$ and
$\varepsilon _{p}^{a}=\widetilde{s}_{p}^{a}/s_{p}^{a}$ are  the dressed
energies (pseudoenergies).
The  free energy  per unit length $f(R,M)$ is
\eq
f(R,M)=-\frac{1}{2\pi }\sum_{a=1}^{N-1}\int_{-\infty }^{\infty }d\theta \
m_{2a} \cosh \theta L_{0-}^{2a}(\theta )~.
\en
In the  ultraviolet limit, $RM \rightarrow 0$,   $f(R,M)$
can be  related  to the central charge of the
corresponding conformal field theory  by 
\eq
f(R,M) \sim -\frac{\pi }{6 R^2 }c_{UV}~.
\en
For the AII sigma model, the natural expectation is
\begin{equation}
c_{UV}=\hbox{Dim}[SU(2N)]-\hbox{Dim}[Sp(N)]= 2N^{2}-N-1~.
\label{Cuv}
\end{equation}
\section{The Y-system}
\label{S3}
\begin{figure}
\begin{center}
\epsfysize=2.5 true in
\hskip 20 true pt
\epsfbox{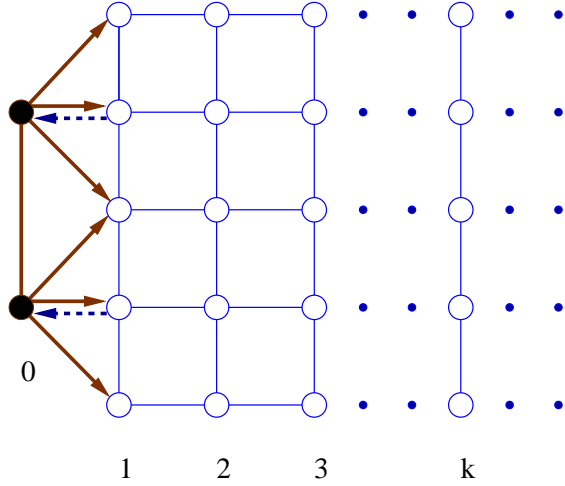}
\end{center}
\caption{The TBA diagram for the $SU(6)/Sp(3)$ sigma model.}
\label{F1}
\end{figure}

A  Y-system is a set  of functional  relations  
among  the   unknown  functions 
\eq
Y_p^a(\theta)=
e^{\varepsilon_p^a (\theta)}~.
\en 
To derive  a Y-system from  the TBA  obtained   in the  previous section
one starts  from   eq. (\ref{Tba0}) in Fourier 
representation, multiply it by $\widehat{K}^{(N)}(2\omega )$ and use 
properties (\ref{Perfr}) and (\ref{INV}) to get
\eq
\frac{\cosh (2\omega )}{\cosh (\omega )}\sum_{b=1}^{N-1}\widehat{K}%
_{ab}^{(N)}(2\omega )\widehat{L}_{0+}^{2b}(\omega
)=\sum_{c=1}^{2N-1}\sum_{q=0}^{\infty }\widehat{K^{\prime }}_{0q}^{(\infty
)}(\omega )\left( 2\cosh (\omega )\delta _{c,2a}+
I_{ac}^{(2N-1)}\right) \widehat{L}_{q-}^{b}(\omega )~,
\label{eq24}
\en
with
\eq
\widehat{K^{\prime }}_{0q}^{(\infty )}(\omega )=\left( 1-\frac{\delta
_{q0}}{%
2\cosh ^{2}(\omega )}\right) \left( \delta _{q0}-\frac{\delta _{q1}}{2\cosh
(\omega )}\right)~.
\en
Equation (\ref{eq24}) can also  be written as 
\bea
\ds{ \frac{\cosh (2\omega )}{\cosh (\omega )}\widehat{L}_{0+}^{2a}(\omega
)-\frac{%
1}{2\cosh (\omega )}\sum_{b=1}^{N-1}I_{ab}^{(N-1)}\widehat{L}_{0+}^{2b}(\omega
)=~~~~~~~~~~~~~~~~~~~~~~~~~~~~~~~~~~~~~} \nn \\
\ds{\frac{\cosh (2\omega )}{\cosh (\omega )}\widehat{L}_{0-}^{2a}(\omega )
-\frac{1}{2\cosh (\omega )}\sum_{b=1}^{2N-1} I_{ab}^{(2N-1)}%
\widehat{L}_{1-}^{b}(\omega )-\widehat{L}_{1-}^{2a}(\omega )}~. 
\eea
By using  $L_{p+}^{a}-L_{p-}^{a}=\varepsilon _{p}^{a}$:   
\eq
2\cosh (2\omega )\widehat{\varepsilon }_{0}^{2a}(\omega )+2\cosh (\omega )%
\widehat{L}_{1-}^{2a}(\omega )=\sum_{b=1}^{N-1}I_{ab}^{(N)}\widehat{L}%
_{0+}^{2b}(\omega )-\sum_{b=1}^{2N-1} I_{ab}^{(2N)}\widehat{L}%
_{1-}^{b}(\omega )~,
\en
or, equivalently
\bea
Y_{0}^{2a}(\theta +2i\pi \Delta )Y_{0}^{2a}(\theta -2i\pi \Delta )
&=& \left( 1+%
\frac{1}{Y_{1}^{2a}(\theta +i\pi \Delta )}\right)^{-1} 
\left( 1+\frac{1}{%
Y_{1}^{2a}(\theta -i\pi \Delta )}\right)^{-1} \nn \\
&\times&  
\prod_{b=1}^{N-1}\left( 1+Y_{0}^{2b}(\theta )\right)
^{I_{ab}^{(N-1)}}\prod_{b=1}^{2N-1}\left( 1+\frac{1}{Y_{1}^{2b}(\theta )}%
\right) ^{-I_{ab}^{(2N-1)}}~.  
\label{Y0}
\eea
($a=1,\dots,N-1$.)
The same procedure  applied to equation (\ref{Tba2}) leads to a  
second set of functional relations of $SU(2N)$ Gross--Neveu type~\cite{Zd,F2}:
\eq
Y_{p}^{c}(\theta +i\pi \Delta )Y_{p}^{c}(\theta -i\pi \Delta
)=\prod_{b=1}^{2N-1}\left( 1+Y_{p}^{b}(\theta )\right)
^{I_{cb}^{(2N-1)}}\prod_{q=0}^{\infty }\left( 1+\frac{1}{Y_{q}^{c}(\theta )}%
\right) ^{-I_{pq}^{(\infty )}}~,  
\label{Y1}
\en
where $p=1,2,\dots,\infty$, $c=1,2,\dots,2N-1$ and by definition  
$Y_{0}^{2i+1}=\infty$.
The  Y-system (\ref{Y0},\ref{Y1}) is  pictorially  represented in 
figure~\ref{F1}.
There is a  one-to-one  correspondence  between the  Y functions and the nodes 
in the diagram. Black nodes correspond
to massive pseudoenergies, while white ones are of  magnonic type;
a link connecting  a node $(p,a)$ and a node $(p',a')$ indicates that when 
$Y^a_p$ appears on  the l.h.s. of equation (\ref{Y1}) (or (\ref{Y0})), 
there is a term containing $Y^{a'}_{p'}$ on the r.h.s. The kind of term 
on the r.h.s. is different for different styles,
horizontal and  vertical  links. 
Arrows indicate one-way directed links.

The ultraviolet central charge can then be written  in terms of the 
sets of numbers  $\{ \cY_p^a \}$ and  $\{ {\cY'}_p^a \}$,
$\theta$-independent solutions of  the 
full Y-system  and of the subsystem  obtained  after eliminating all the
 massive nodes:
\begin{equation}
c_{UV}=\frac{6}{\pi ^{2}}\left[ \sum_{all\hspace{0.2cm}nodes} L \left(
\frac{1}{1+\cY_{p}^{a}}\right)%
-\sum_{magnonic\hspace{0.2cm}nodes}%
L \left( \frac{1}{1+{\cY'}_{p}^{a}}\right) %
\right]~, 
 \label{c}
\end{equation}%
where $L$ is the Rogers  dilogarithm function. 
\section{Truncation of the Y-system and its ultraviolet analysis}
\label{Sn}
One of the main results  of
\cite{KN} is an exact conjecture for   sum-rules of the type (\ref{c}) 
for   Y-systems related to  the classical Lie algebras
\eq
\frac{6}{\pi ^{2}}\sum_{all\hspace{0.2cm}nodes}L\left( \frac{%
\cY_{p}^{{a}{\bf [KN]}}}{1+\cY_{p}^{{a}{\bf [KN]}}}\right) =\frac{l~\hbox{Dim}[G]}{l+g^{\vee}}-\hbox{rank}[G]~.
\en

Notice that the first term in the r.h.s.  is the central charge of a WZW
model for the algebra $G$ at level $l$ 
($g^{\vee}$ is the dual Coxeter number). 
It is now easy  to check  that the  Y-system (\ref{Y0},\ref{Y1}) 
becomes  equivalent to the  
$C_{k+1}$-related  system of \cite{KN} at  $l=N$, provided  
the infinite TBA diagram is truncated  in the horizontal 
direction up to the node $k$ (see figure~\ref{F1}) and the following identifications are made 
\eq
Y_p^a(\theta) = 1/Y_p^{{a}{\bf [KN]}}(\theta)~,~~~ 
g^{\vee}   = k+2~,~~~ 
\hbox{rank}[G]  = k+1~, 
\en
and hence
\eq
c_1=\frac{6}{\pi ^{2}}  \sum_{all~nodes} L \left(
\frac{1}{1+\cY_{p}^{a}}\right)
=
\frac{N(k+1)(2(k+1)+1)}{N+k+2}-k-1~.
\label{c1}
\en
After   the elimination of  the massive nodes instead the correspondence 
is  with a system of $A_{k}$ type at level  $l=2N$. 
The second contribution  to  (\ref{c}) is 
\eq
c_2=\frac{6}{\pi ^{2}} \sum_{magnonic~nodes}
L \left( \frac{1}{1+{\cY'}_{p}^{a}}\right) %
=  
\frac{2Nk(k+2)}{2N+k+1}-k~.
\label{c2}
\en
Notice now that
\eq
c_1-c_2 = c[SU(2N)_{k+1}]-c[Sp(N)_{k+1}]~,
\en
is equal to  the difference in the central charges  
of the  $SU(2N)_{k+1}$ and $Sp(N)_{k+1}$  WZW  models. 
It is then  natural to  conjecture that the level $k$ truncated 
system describes an integrable  perturbation of the conformal  model 
$SU(2N)_{k+1}/Sp(N)_{k+1}$.
The perturbing operator is easily identified by noticing that  
the solutions of the Y-systems fulfil the periodicity property~\cite{Zam2,T} 
\eq
Y_p^a \left(\theta+i \pi {N+k+2 \over 2N} \right)= Y_p^{2N-a}(\theta)~, 
\en 
then the argument of \cite{Zam2} leads to the conformal dimensions
\eq
h=\bar{h}=1- {2N \over  N+k+2}~,
\en
suggesting  that the models in question are most certainly  perturbed
by the operator   $(1,adj)$.

\noindent
On the other hand, in  the $k \rightarrow \infty$ limit   
$h=\bar{h} \rightarrow 1$ and   
\eq
c_{UV}=\lim_{k\rightarrow \infty }(c_{1}-c_{2})=
2N^{2}-N-1~,  
\label{cuv}
\end{equation}%
which is  the expected  result ~(\ref{Cuv}) 
for  the  AII nonlinear sigma model.

\section{Conclusions}
\label{S4}

The main purpose of this paper is to show  that the $SU(2N)$  Gross--Neveu S-matrix,  with a CDD factor
selecting  only even fundamental highest weight multiplets, correctly
describes the ultraviolet  limit of the sigma models on the AII
symmetric spaces. Considering also the earlier result   of  \cite{My}, where
the  $(\mu _{2}, \mu_{2})$  scattering  amplitude 
was checked  by comparing the  free energy
in strong external field    with  perturbation theory, we  can 
confidently confirm the correctness  
of the  proposal~\cite{My}.
 
Another result  of the paper is the mapping, through the Y-system, 
between our models and  particular limits of  systems proposed  by Kuniba and 
Nakanishi in~\cite{KN}. This   led us to the interesting conjecture that the 
truncated TBA equations describe  the coset 
$SU(2N)_{k+1}/Sp(N)_{k+1}$ models  perturbed by the operator $(1,adj)$.
As a follow-up project it  would be interesting  to check the integrability 
of these perturbed conformal field theories, 
to propose the exact  S-matrix and
to rigorously derive the set of thermodynamic equations.

A more general conclusion is again related to 
the interesting fact  that  the finite-size physics  
of these  more exotic  models  is  again described by set of equations 
of the kinds~\cite{KN,RTV}. It would be nice to understand this 
correspondence at  a deeper level and to find  the  complete classification of 
the thermodynamic equations for integrable  sigma models on symmetric spaces.

\bigskip
{\bf Acknowledgements}

\noindent
We are grateful to 
Patrick Dorey and Sergei Lukyanov  for useful discussions.
AB thanks  the Mathematics Department of Durham University  for 
hospitality  while some of this work was in progress
and RT thanks the EPSRC for an Advanced Fellowship.

\section{Appendix}
The definition of   Fourier transform is
\eq
f(\theta )=g \int_{-\infty }^{\infty }\frac{d\omega }{2\pi^2 }
e^{i\frac{g}{\pi }\theta
\omega }\widehat{f}(\omega )~,~~~~\ \widehat{f}(\omega
)=\int_{-\infty }^{\infty } d\theta~ e^{-i\frac{g}{\pi }\theta \omega }f(\theta
)~,
\en
where $g=2N$ is the dual Coxeter number 
of $SU(2N)$. The kernels 
$\widehat{A}_{ab}^{(L)}(\omega )$ and $\widehat{\Psi}_{2a2b}$ are
\eq
\widehat{A}_{ab}^{(L)}(\omega ) = 2\cosh (\omega )\frac{\sinh (\omega \min
(a,b))\sinh (\omega (L-\max (a,b)))}{\sinh (\omega L)\sinh (\omega )}~,
\en
\eq
\widehat{\Psi}_{2a2b}(\omega ) = e^{-|\omega |}
\frac{\sinh(2\omega \min (a,b))\sinh 
(2\omega (N-\max (a,b)))}{\sinh (2\omega N)\sinh (\omega )}~,
\en
\eq
\widehat{K}_{ab}^{(L)}(\omega ) =
\left( \widehat{A}_{ab}^{(L)}(\omega )\right)^{-1}=
\delta _{ab}-\frac{I_{ab}^{(L-1)}}{2\cosh (\omega )}~,  
\en
\eq
\widehat{a}_{p}(\omega ) = e^{-p|\omega |}~.   
\en
$I^{(L-1)}=2\delta _{ab}-C_{ab}^{A_{L-1}}$  and  $C^{A_{L-1}}$ are, 
respectively,  
the incidence and Cartan matrices  of the algebra $A_{L-1}$. 
The following relations are also  useful
\eq
\sum_{b=1}^{\infty }\widehat{K}_{ab}^{(\infty )}(\omega )\widehat{a}%
_{b}(\omega )=\frac{\delta _{a1}}{2\cosh (\omega )}~,
\en
\begin{equation}
\sum_{b=1}^{N-1}\widehat{K}_{ab}^{(N)}(2\omega )m_{2b}\widehat{\cosh \theta
}%
=0~,  
\label{Perfr}
\end{equation}%
\eq
\sum_{b=1}^{N-1}\widehat{K}_{ab}^{(N)}(2\omega )\widehat{A}%
_{2b,c}^{(2N)}(\omega )=\frac{\cosh (\omega )}{\cosh (2\omega )}\left(
2\cosh (\omega )\delta _{c,2a}+I_{2a,c}^{(2N)}\right)~.  
\label{INV}
\en
In equation (\ref{Perfr}) $m_{2b}=M\sin \left( \frac{\pi b}{N}\right) $ is the Perron-Frobenius eigenvector of the $A_{N-1}$ Cartan matrix.

\end{document}